# Photocarrier injection and the *I-V* characteristics of $La_{0.8}Sr_{0.2}MnO_3/SrTiO_3$:Nb heterojunctions

Takaki Muramatsu[*], Yuji Muraoka, Zenji Hiroi

*Institute for Solid States Physics, University of Tokyo, Kashiwa, Chiba 277 8581, Japan*

Abstract

Oxide heterojunctions made of p-type $L_{0.8}Sr_{0.2}MnO_3$ (LSMO) and niobium-doped n-type $SrTiO_3$ (STO:Nb) have been fabricated by the pulsed laser deposition (PLD) technique and are characterized under UV light irradiation by measuring the current-voltage, photovoltaic properties and the junction capacitance. It is shown that the heterojunctions work as an efficient UV photodiode, in which photogenerated holes in the STO:Nb substrate are injected to the LSMO film. The maximum surface hole density $Q/e$ and external quantum efficiency $\gamma$ are estimated to be $8.3 \times 10^{12}$ cm$^{-2}$ and 11 % at room temperature, respectively. They are improved significantly in a p-i-n junction of LSMO/STO/STO:Nb, where $Q/e$ and $\gamma$ are $3.0 \times 10^{13}$ cm$^{-2}$ and 27 %, respectively.

* Corresponding author. Tel.: +81-4-7136-3447; fax: +81-4-7136-3446.

E-mail address: muramatu@issp.u-tokyo.ac.jp

Strongly correlated electron systems in transition metal oxides (TMOs) have been investigated for a long time because of their interesting properties full of variety, including the Mott transition, high $T_c$ superconductivity, colossal magnetoresistance and so on. Especially, in Mn perovskite oxides, complex interplay between magnetic and electronic properties as a function of doped hole concentration attract a great deal of interest [1]. The hole concentration is usually controlled by the chemical substitution of constituent elements as in $La_{1-x}Sr_xMnO_3$. Recently, Muraoka *et al.* discovered an alternative way of hole doping which is the photocarrier injection (PCI) method in oxide heterostructures under UV light irradiation [2,3]. This method is expected to make it possible to cleanly dope TMOs or organics with holes and has been applied to vanadium oxides, manganites, organics and copper oxide superconductors [4-6].

PCI has been carried out by irradiating with UV light a heterostructure made of a p-type TMO film and an n-type $SrTiO_3$ (STO) or $TiO_2$ substrate. The heterostructure is considered to be a p-n heterojunction as in conventional semiconductor diodes. Typically, a TMO has a bandgap of about 1.0 eV, while that of the titanates is much larger, 3.2 eV for $SrTiO_3$. When a heterojunction between the two types of compounds is formed, a potential barrier, which height is determined by the difference in the work function between them, is built in the conduction band of the titanate substrates near the boundary. On the other hand, the top of the valance band is always higher in the TMO films than in the substrates. Under UV light irradiation, hole and electron pairs are generated in the substrate, and only holes can be injected to the film. Therefore, it is considered that holes are doped cleanly into the TMO film without any chemical substitutions, and that the hole density is controlled by the light intensity. In this letter, we apply the PCI method to Mn perovskite oxides and investigate the efficiency of hole doping. It is also shown that the present oxide heterojunction

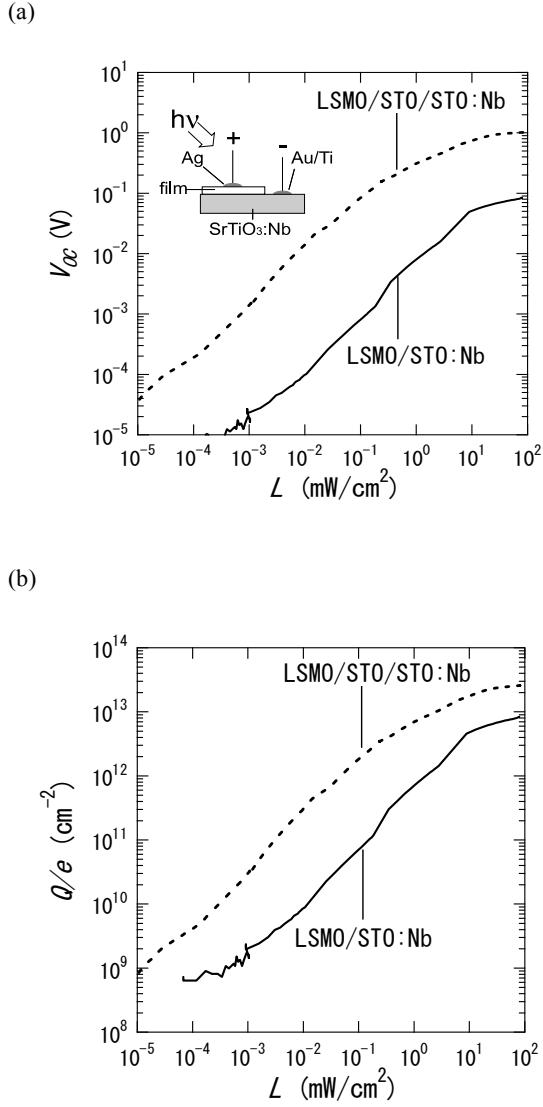

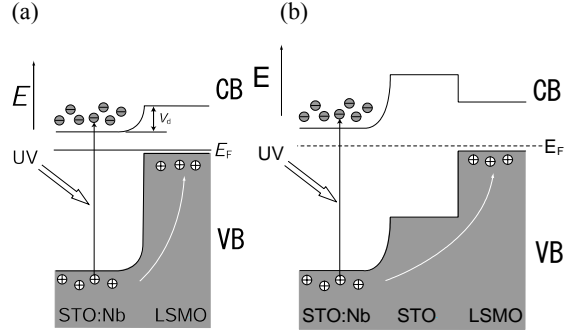

Fig. 2 Schematic band diagrams for (a) La$_{0.8}$Sr$_{0.2}$MnO$_3$/SrTiO$_3$:Nb and (b) La$_{0.8}$Sr$_{0.2}$MnO$_3$/SrTiO$_3$/SrTiO$_3$:Nb heterojunctions. CB and VB refer to the conduction and valence bands, respectively, and V$_d$ the diffusion voltage.

Fig. 1. UV light irradiance dependence of open-circuit photovoltage $V_{oc}$ for the La$_{0.8}$Sr$_{0.2}$MnO$_3$/SrTiO$_3$:Nb (solid line) and La$_{0.8}$Sr$_{0.2}$MnO$_3$/SrTiO$_3$/SrTiO$_3$:Nb junctions (doted line). Inset shows the schematic alignment of electrodes.
(b) UV light irradiance dependence of the surface hole density $Q/e$ for the La$_{0.8}$Sr$_{0.2}$MnO$_3$/SrTiO$_3$:Nb (solid line) and La$_{0.8}$Sr$_{0.2}$MnO$_3$/SrTiO$_3$/SrTiO$_3$:Nb junctions (doted line).

exhibits a notable feature as a UV photodiode.

Two heterojunctions were fabricated by utilizing the pulsed laser deposition (PLD) technique. One is a p-n heterojunction in which a La$_{0.8}$Sr$_{0.2}$MnO$_3$ (LSMO) film is grown on a 0.05 wt% Nb-doped SrTiO$_3$ (100) single crystal at a substrate temperature of 750-800 ℃ in an oxygen partial pressure of 0.1 torr. A sintered La$_{0.8}$Sr$_{0.2}$MnO$_3$ pellet was used as a target. The film thickness is about 40 nm. The other is a p-i-n heterojunction with a non-doped STO layer of 20 nm thick inserted between a STO:Nb substrate and a LSMO film. The deposition of the non-doped STO layer was performed by PLD from a STO pellet. The $I$-$V$ characteristics and open-circuit voltage $V_{oc}$ were measured in the dark and under UV light irradiation. A UV light with $\lambda$ = 30~40 nm from a Xe Lamp was used. A wide range of UV light intensity from 10$^{-5}$ to 10$^2$ mW/cm$^2$ was obtained by using variable ND filters. A silver paste was used as an electrode for the LSMO film, and Au/Ti contacts were deposited on the STO:Nb substrates under high vacuum condition. The capacitance of the junctions was measured by using the HP-4282 precision LCR meter working at a frequency of 20 Hz and the amplitude of 20 mV.

A positive photovoltage to the film has been observed for the LSMO/STO:Nb heterojunction under UV light irradiation. It is shown in Fig. 1 as a function of UV light intensity $L$. The photovoltage increases almost linearly with light intensity in a wide UV light intensity range of 10$^{-4}$ < $L$ < 10 mW/cm$^2$ and then tends to saturate to the maximum value of 0.09 V at 80 mW/cm$^2$. This value is considerably smaller than those found in VO$_2$/TiO$_2$:Nb (0.4 V) and YBa$_2$Cu$_3$O$_{7-\delta}$/STO:Nb (0.8 V). The linearity at low irradiance is understood by the rigid band model shown in Fig. 2, while the saturation at high irradiance implies that the photovoltage becomes comparable to the height of the potential barrier for electrons. Therefore, the maximum photovoltage of 0.09 V corresponds to the effective height of the potential barrier at room temperature.

In order to determine the height of the potential barrier or the diffusion voltage $V_d$, we measured capacitance $C$ of LSMO/STO:Nb. Assuming that the depletion layer is formed only in the STO:Nb substrate, junction capacitance can be expressed as $C^2 = eN_D\varepsilon_0\varepsilon_s/2(V_d-V)$, where $N_D$, $\varepsilon_0$ and $\varepsilon_s$ are donor (Niobium) concentration, vacuum and relative dielectric constants, respectively. Thus, one expects a linear relation between $C^{-2}$ and $V$ Fig. 3 shows the $C^{-2}$ versus $V$ plot measured at room temperature and 100 K. A linear dependence is seen only at low temperature, as a leak current becomes negligible at low temperature. Using the low temperature data, the $V_d$ is determined to be 0.45 V from the voltage where the linear extrapolation intercepts the $V$ axis. Moreover, the $\varepsilon_s$ is estimated to be 340 from the slope of the $C^{-2}$-$V$ line, which is nearly equal to the previously reported value for a STO single crystal ($\varepsilon_s$ = 330).

We have estimated the injected surface hole density $Q/e$ by integrating the capacitance equation $dQ = C_d V_{oc}$, because this type of junction is regarded as a condenser [3]. At low irradiance the $Q/e$ is proportional to the light irradiance and begins to saturate over $L$ = 10 mW/cm$^2$ as shown in Fig. 1(b). The maximum surface hole density is $8.3 \times 10^{13}$ cm$^{-2}$ at $L$ = 90 mW/cm$^2$. Assuming uniform distribution of holes in the 10-nm-thick LSMO film, the maximum hole concentration by PCI would be 0.06 % per Mn. Apparently this value is too small to expect a notable change in the properties of LSMO film. However, since injected holes can be accumulated near the boundary as discussed in the case of YBCO/STO:Nb [7], actual hole density may be larger.

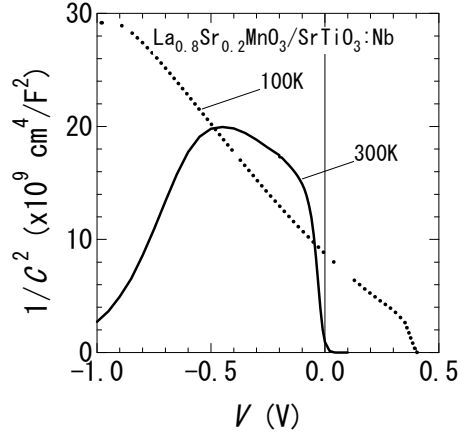

Fig. 3. Bias voltage dependence of $1/C^2$ for the La$_{0.8}$Sr$_{0.2}$MnO$_3$/SrTiO$_3$:Nb junction in the dark with an ac frequency of 20 Hz and the amplitude of 20 mV.

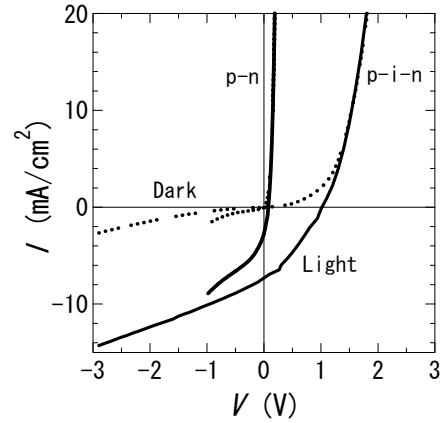

Fig. 4. I-V characteristics of the La$_{0.8}$Sr$_{0.2}$MnO$_3$/SrTiO$_3$:Nb (p-n) and La$_{0.8}$Sr$_{0.2}$MnO$_3$/SrTiO$_3$/SrTiO$_3$:Nb (p-i-n) junctions in the dark and under UV light irradiation.

One of the reasons for the low $Q/e$ in LSMO/STO:Nb is the small $V_d$, which comes from the small difference in the work function between LSMO and STO:Nb. In order to enhance the efficiency of PCI in the Mn perovskites, another heterostructure was fabricated, where a non-doped SrTiO$_3$ layer with 20 nm thickness was sandwiched between a LSMO film and a STO:Nb substrate. This junction is considered as a p-i-n junction, which is often used in a solar cell for the purpose of improving the quantum efficiency. A possible band diagram for this p-i-n (LSMO/STO/STO:Nb) junction is shown in Fig. 2. In this structure the insulating STO layer should separate photogenerated hole and electrons spatially, and thus gives rise to a large potential barrier for electrons. The height of the barrier is expected to be approximately half of the bandgap of STO, that is, $V_d \sim 1.6$ eV. As a result, injected holes would stay in the LSMO film with a longer life time. The $V_{oc}$ of the LSMO/STO/STO:Nb junction is shown in Fig. 1. Apparently, the $V_{oc}$ is enhanced compared with case of LSMO/STO:Nb by a factor of 100 at low irradiance and 10 at high irradiance. The saturation value at room temperature is about 1.0 V, ten times larger than that of the p-n junction. Although the inserted STO layer may reduce the junction capacitance, the maximum surface hole density is estimated to be $3.0 \times 10^{13}$ cm$^{-2}$, about four times larger than in the p-n junction. Therefore, the efficiency of PCI is improved significantly in the p-i-n junction, though the resulting hole density may be still too small to control the transport of the LSMO film. For further improvement, the use of a substrate material with a larger dielectric constant would be critical.

The present junctions exhibit photodiode features as shown in the current-voltage ($I$-$V$) characteristics measured in the dark and under UV light irradiation with $L$ = 79 mW/cm$^2$ at room temperature in Fig. 4. These heterostructures show rectifying behavior in the dark with a leakage current density of 1.7 mA/cm$^2$ and 0.5 mA/cm$^2$ at a bias voltage of -1.0 V for the p-n and p-i-n junctions, respectively. Under UV light illumination the $I$-$V$ curves shift downward as generally seen in conventional semiconductor photodiodes. The short circuit current $I_{sc}$ is about 2.7 mA/cm$^2$ and 7.3 mA/cm$^2$, at $L \sim 80$ mW/cm$^2$ and the external quantum efficiency is $\gamma$ = 11% and 28 %, respectively. Comparing the two junctions, it is apparent that the insertion of the insulating STO layer gives rise to the enhancement of $I_{sc}$ and $V_d$, resulting in the larger quantum efficiency.

In conclusion, we fabricated the high quality heterojunctions of $La_{0.8}Sr_{0.2}MnO_3/SrTiO_3$:Nb and found the apparent photocarrier injection to the $La_{0.8}Sr_{0.2}MnO_3$ film under UV light irradiation. The maximum doped hole density is $8 \times 10^{12}$ cm$^{-2}$ which is not large enough to control the transport properties of the film. The junction works as a UV photodiode with an external quantum efficiency of 11 %. It is revealed that the efficiency can be improved markedly by the formation of a p-i-n junction of $La_{0.8}Sr_{0.2}MnO_3/SrTiO_3/SrTiO_3$:Nb, where the injected hole density and quantum efficiency are $3.0 \times 10^{13}$ cm$^{-2}$ and 28 %, respectively.

We thank T Yamauchi for his help in electrical measurement and J. Yamaura for helpful discussion. This work is supported by a Grant-in-Aid for Creative Scientific Research (IJNP0201) and a Grant-in-Aid for Scientific Research (14750549) given by the Ministry of Education, Culture, Sports, Science and Technology.

Reference

[1] A. Urushibara, Y. Moritomo, T. Arima, A. Asamitsu, G. Kida and Y. Tokura, Phys. Rev. B 51 (1995) 14103.

[2] Y. Muraoka, T. Yamauchi, Y. Udeda and Z. Hiroi, J. Phys.:Condens. Matter 14 (2002) L757.

[3] Y. Muraoka and Z. Hiroi, J. Phys. Soc. Jpn. 72 (2003) 781

[4] J. Yamaura, Y. Muraoka, T. Yamauchi, T. Muramatsu, and Z. Hiroi, Appl. Phys. Lett. 83 (2003) 2097

[5] Y. Muraoka, T. Yamauchi, T. Muramatsu, J. Yamaura and Z. Hiroi, J. Phys. Soc. Jpn. 72 (2003) 3049.

[6] T. Muramatsu, Y. Muraoka, T. Yamauchi, J. Yamaura and Z. Hiroi, J. Magn. Magn. Mater. to be published.

[7] Y. Muraoka, T. Yamauchi, T. Muramatsu, J. Yamaura and Z. Hiroi, submitted to Appl. Phys. Lett.